\documentclass[cernpreprint, atlasdraft=false, UKenglish, texmf, orcidlogo]{atlasdoc}
 
\usepackage{atlaspackage}
\usepackage{atlasbiblatex}
 
\usepackage{atlasphysics}

\graphicspath{{logos/}{figures/}}
 
\usepackage{ANA-HIGG-2019-16-PAPER-defs}

\AtlasTitle{Measurement of the Higgs boson mass with
$H \rightarrow \gamma\gamma$ decays in 140~fb$^{-1}$ of $\sqrt{s}=13$~TeV $pp$
collisions with the ATLAS detector}

\AtlasAbstract{
The mass of the Higgs boson is measured in the $H\to\gamma\gamma$
decay channel, exploiting the high resolution of
the invariant mass of photon pairs reconstructed from the decays of Higgs
bosons produced in proton--proton collisions at a centre-of-mass
energy $\sqrt{s}=13$~TeV.
The dataset was collected between 2015 and 2018 by the ATLAS detector
at the Large Hadron Collider,
and corresponds to an integrated luminosity of 140~fb$^{-1}$.
The measured value of the Higgs boson mass is
$125.17 \pm 0.11~\mathrm{(stat.)} \pm 0.09~\mathrm{(syst.)}$~GeV
and is based on an improved energy scale calibration for photons, whose
impact on the measurement is about four times smaller than
in the previous publication.
A combination with the corresponding measurement using 7
and 8~TeV $pp$ collision ATLAS data results in a Higgs boson
mass measurement of
$125.22 \pm 0.11~\mathrm{(stat.)} \pm 0.09~\mathrm{(syst.)}$~GeV.
With an uncertainty of 1.1 per mille, this is currently the most
precise measurement of the mass of the Higgs boson from a single decay
channel.
}

\AtlasRefCode{HIGG-2019-16}
 
\PreprintIdNumber{CERN-EP-2023-160}

\AtlasJournalRef{Phys. Lett. B 847 (2023) 138315}
\AtlasDOI{10.1016/j.physletb.2023.138315}

\hypersetup{pdftitle={ATLAS document},pdfauthor={The ATLAS Collaboration}}
 
\begin{document}
 
\maketitle
 
\tableofcontents

% The next lines are included from the .//sections/01-intro.tex input file
\section{Introduction}
 
After the ATLAS and CMS collaborations at the Large Hadron Collider
(LHC) discovered in 2012~\cite{HIGG-2012-27,CMS-HIG-12-028}
a particle $H$ with properties consistent with those
of the Higgs boson in the Standard Model (SM) of particle physics,
the precise determination of its mass became
one of the primary goals of their physics programmes.
The Higgs boson mass $m_H$ is a fundamental parameter of the
SM and the only unknown parameter of the scalar sector
of the Standard Model prior to the Higgs boson discovery.
Its measurement is of paramount importance for several reasons.
Firstly, its value determines the Higgs boson production rates
and decay branching ratios~\cite{deFlorian:2016spz}. It is also the value
assumed by the experimental collaborations when estimating the acceptances,
efficiencies and signal models used in their analyses and to report
their measured rates.
Knowledge of the Higgs boson mass is therefore mandatory for
a coherent test of its coupling structure.
Secondly, the Higgs boson mass is one of the inputs in
global fits to the measurements of electroweak
observables~\cite{Haller:2018nnx}.
Knowing its value therefore plays a key role in verifying the internal
consistency of the SM, especially through the interplay between the masses
of the top quark and the $W$ and Higgs bosons.
Finally, the stability of the electroweak vacuum, and thus the fate
of our universe, depends on the value of the Higgs boson
mass~\cite{Sher:1988mj}. By measuring $m_H$, one can infer
whether the universe is in a global,
and thus stable, minimum-energy state of the Higgs field potential,
or in a local metastable one, from which it could
decay to the ground state in the future~\cite{Degrassi:2012ry}.
 
Measurements of the Higgs boson mass were performed separately
by the ATLAS~\cite{HIGG-2013-12}
and CMS~\cite{CMS-HIG-14-009} collaborations using
Higgs boson decays to the high-resolution
four-lepton ($4\ell$, $\ell=e,\mu$) and diphoton
($\gamma\gamma$) final states reconstructed during the first
data-taking phase of the LHC (\RunOne).
The data consisted of $25$~\ifb~of proton--proton ($pp$) collisions
recorded at centre-of-mass energies $\sqrt{s}=7$ and 8~\TeV\ in
2011 and 2012.
The combination of the ATLAS and CMS results led to a measurement of
the Higgs boson mass with an uncertainty of 0.19\%,
$\mH=125.09\pm 0.24$~\GeV~\cite{HIGG-2014-14}.
 
Updated measurements of the Higgs boson mass were performed by both
experiments using $pp$ collisions collected at $\sqrt{s} = 13$~\TeV\
between 2015 and 2018 during the second data-taking phase of
the LHC (\RunTwo).
Using both \Hllll and \Hyy decays selected in a partial \RunTwo dataset
(36~\ifb\ of $pp$ collisions recorded before 2017),
the ATLAS Collaboration measured
$m_H = 124.86 \pm 0.27$~\GeV~\cite{HIGG-2016-33}.
Combined with the ATLAS \RunOne results from Ref.~\cite{HIGG-2013-12},
this study led to a Higgs boson mass measurement
of $\mH = 124.97 \pm 0.24$~\GeV.
Using a dataset of equivalent size and both the four-lepton and diphoton
final states, the CMS Collaboration found
$m_H = 125.46 \pm 0.16$~\GeV~\cite{CMS-HIG-19-004},
whose combination with the \RunOne results from Ref.~\cite{CMS-HIG-14-009}
led to the most precise determination of \mH to date, with a 0.11\%
uncertainty: $\mH = 125.38 \pm 0.14$~\GeV.
Recently, the ATLAS Collaboration released an updated measurement of
\mH using \Hllll decays in the full \RunTwo dataset~\cite{ATLAS:2022net},
consisting of 140~\ifb of $pp$ collisions.
The result, $\mH = 124.99 \pm 0.19$~\GeV,
combined with that of the corresponding analysis using \RunOne data,
yields a single-channel Higgs boson mass measurement with an
uncertainty of 0.14\%, $\mH = 124.94 \pm 0.18$~\GeV.
 
This Letter reports
a measurement of the Higgs boson mass in the diphoton
channel using the full \RunTwo dataset.
Compared with that in Ref.~\cite{HIGG-2016-33}, the analysis presented
here profits from a larger data sample, a new photon
reconstruction algorithm with better energy resolution~\cite{EGAM-2018-01},
an improved estimation of the photon energy scale with reduced
uncertainties~\cite{newegcalib},
and an optimised event classification strategy.
The selected events are required to contain two energetic photons
fulfilling strict identification and isolation criteria.
The invariant mass (\myy) distribution of the selected photon
pairs exhibits a peak near \mH, arising from resonant Higgs boson
decays, over a smoothly falling distribution from background events
mainly due to non-resonant diphoton production.
The Higgs boson mass is determined from the position of the peak
in data through a profile-likelihood fit to the $\myy$ distribution.
Simulated signal and background event samples are used to optimise the
analysis criteria, to define the signal and background $\myy$ models
used in the fit, and to estimate the impact of the systematic uncertainties
in \mH.
To increase the sensitivity of the measurement, the selected
events are classified into mutually exclusive categories
with different diphoton invariant mass resolutions
and signal-to-background ratios which are analysed simultaneously.
Finally, a combination with the ATLAS \RunOne measurement~\cite{HIGG-2014-14}
is performed.
 
The rest of the Letter is organised as follows.
The ATLAS detector is described briefly in Section~\ref{sec:detector}.
The data and simulated event samples used in the analysis are summarised in
Section~\ref{sec:samples}.
The photon reconstruction and the event selection and classification are
discussed in Section~\ref{sec:selection}.
The statistical tools used in the measurement and the methods used to
assess the systematic uncertainties are presented in
Sections~\ref{sec:methodology} and~\ref{sec:syst},
leading to the results described in Section~\ref{sec:results}.
The conclusions of this study are summarised in Section~\ref{sec:conclusion}.

% End of text imported from the .//sections/01-intro.tex input file

% The next lines are included from the .//sections/02-detector.tex input file
\section{The ATLAS detector}
\label{sec:detector}
 
\newcommand{\AtlasCoordFootnote}{
ATLAS uses a right-handed coordinate system with its origin at the nominal
interaction point (IP) in the centre of the detector and the \(z\)-axis
along the beam pipe.
The \(x\)-axis points from the IP to the centre of the LHC ring,
and the \(y\)-axis points upwards.
Cylindrical coordinates \((r,\phi)\) are used in the transverse plane,
\(\phi\) being the azimuthal angle around the \(z\)-axis.
The pseudorapidity is defined in terms of the polar angle \(\theta\) as
\(\eta = -\ln \tan(\theta/2)\).
Angular distance is measured in units of \(\Delta R \equiv \sqrt{(\Delta\eta)^{2} + (\Delta\phi)^{2}}\).
The transverse energy is defined as $\et = E\sin(\theta)$.
}
 
The ATLAS experiment~\cite{PERF-2007-01} at the LHC is a multipurpose particle
detector with a forward--backward symmetric cylindrical geometry and a near
\(4\pi\) coverage in  solid angle.\footnote{\AtlasCoordFootnote}
It consists of an inner tracking detector surrounded by a thin superconducting
solenoid providing a \SI{2}{\tesla} axial magnetic field, electromagnetic and
hadron calorimeters, and a muon spectrometer.
The inner tracking detector covers the pseudorapidity range \(|\eta| < 2.5\).
It consists of silicon pixel, silicon microstrip, and transition radiation
tracking detectors.
Lead/liquid-argon (LAr) sampling calorimeters provide electromagnetic (EM)
energy measurements with high granularity.
A steel/scintillator-tile hadron calorimeter covers the central pseudorapidity
range (\(|\eta| < 1.7\)).
The endcap and forward regions are instrumented with LAr calorimeters
for both the EM and hadronic energy measurements up to \(|\eta| = 4.9\).
The muon spectrometer surrounds the calorimeters and is based on
three large superconducting air-core toroidal magnets with eight coils each.
The field integral of the toroids ranges between \num{2.0} and
\SI{6.0}{\tesla\metre} across most of the detector.
The muon spectrometer includes a system of precision chambers for tracking
and fast detectors for triggering.
 
A two-level trigger system is used to select events.
The first-level trigger is implemented in hardware and uses a subset of the
detector information to accept events at a rate below \SI{100}{\kHz}.
This is followed by a software-based trigger that
reduces the accepted event rate to \SI{1}{\kHz} on average
depending on the data-taking conditions.
 
An extensive software suite~\cite{ATL-SOFT-PUB-2021-001} is used
in data simulation, in the reconstruction and analysis of real and simulated data,
in detector operations, and in the trigger and data acquisition
systems of the experiment.

% End of text imported from the .//sections/02-detector.tex input file

% The next lines are included from the .//sections/03-samples.tex input file
\section{Data and simulation samples}
\label{sec:samples}
 
\subsection{Data}
The measurement uses the full $pp$ collision dataset collected
at $\sqrt{s}=13$~\TeV\ by the ATLAS detector
during \RunTwo of the LHC.
Events were recorded using unprescaled diphoton and single-photon triggers~\cite{TRIG-2018-05}.
The photon transverse momentum thresholds were 35~\GeV\ and 25~\GeV\ for
the diphoton triggers throughout the whole of \RunTwo,
and 120 (140)~\GeV\ for single-photon triggers in 2015--2016 (2017--2018).
Shower-shape requirements looser than those used in the offline analysis
were applied to the photon candidates at the trigger level.
The integrated luminosity of the dataset after trigger and data-quality
requirements~\cite{DAPR-2018-01} is
$140.1\pm 1.2$~\ifb~\cite{DAPR-2021-01,Avoni_2018}.
The efficiency of the trigger system for signal events passing the full
selection is close to 100\%~\cite{TRIG-2018-05}.
 
\subsection{Simulation samples}
Monte Carlo (MC) simulated event samples of Higgs bosons produced
by $pp$ collisions and decaying into diphotons,
as well as samples of the main background
processes for the same final state,
were produced with the set-up described in Ref.~\cite{ATLAS:2022tnm}.
Simulated hard-scattering events were overlaid with
simulated inelastic proton--proton events generated with
\PYTHIA[8.1]~\cite{Sjostrand:2007gs},
to model the effect of multiple \enquote{\pileup{}} interactions in the same
and neighbouring bunch crossings.
 
Signal samples were produced for the main Higgs boson production modes:
gluon--gluon fusion (ggF), vector-boson fusion (VBF), and associated production
with a vector boson ($VH$, $V=W,Z$), a top-quark pair ($t\bar{t}H$), a
bottom-quark pair ($b\bar{b}H$) or a single top quark ($tH$).
Signal event samples were produced with either the
\POWHEGBOX~\cite{Alioli:2010xd} or
(for $tH$ only) \MGNLO~\cite{Alwall:2014hca}
event generator, using matrix element calculations at the highest available
order of accuracy in the strong coupling constant \alphas:
next-to-next-to-leading order (NNLO) for ggF,
next-to-leading order (NLO) for VBF, $WH$, $q\bar{q}\to ZH$,
$t\bar{t}H$, $b\bar{b}H$ and $tH$, and leading order (LO) for $gg\to ZH$.
The event generators were interfaced to
\Pythia[8.2]~\cite{Sjostrand:2014zea}
for the modelling of the parton shower and the underlying event.
In the analysis the samples are normalised to the integrated luminosity
of the data, using state-of-the-art Standard Model calculations for the
Higgs boson production  cross-sections and branching ratios at the
hypothesised Higgs boson mass~\cite{deFlorian:2016spz}.
The generated signal samples
were passed through a detailed simulation of the response
of the ATLAS detector~\cite{SOFT-2010-01} based on
\GEANT~\cite{Agostinelli:2002hh}.
 
The nominal signal samples were generated assuming a Higgs boson mass of
125~\GeV.
The Higgs boson width $\Gamma$ in all signal samples was set to the SM
prediction for $m_H=125$~\GeV, $\Gamma = 4.07$~\MeV,
which is much narrower than the experimental resolution.
Systematic uncertainties related to the modelling of the parton shower
are studied with alternative samples produced with the same
matrix-element
generator as used for the nominal ones but with the \HERWIG[7.1.3] parton shower
algorithm~\cite{Bellm:2017jjp}.
The parameterisations of the expected signal yields and diphoton invariant mass
distributions as a function of the Higgs boson mass are obtained by
interpolation between results from signal samples with $m_H$ set to
110, 122, 123, 124, 125, 126, 127, 130 or 140~\GeV, as
described in Section~\ref{sec:methodology}.
The same set-up as that for the nominal samples was used.
 
Background events from non-resonant $pp\to\gamma\gamma + n$~parton
$(n \geq 0)$ production were also simulated,
using the \Sherpa[2.2.4] event generator~\cite{Bothmann:2019yzt}
with NLO-accurate matrix elements for up to one parton,
and leading-order (LO) accurate matrix elements for up to three partons.
Due to its large size, the $pp\to\gamma\gamma$ sample
was processed by a fast simulation of the
ATLAS detector~\cite{ATL-PHYS-PUB-2010-013}, based on a parameterisation
of the response of the calorimeter.
Since the diphoton background is estimated from the sidebands in the
diphoton invariant mass distribution in data,
the background simulation is only used to select the background model
and the systematic uncertainty associated with this choice, and for this
the fast simulation was found to be sufficiently accurate.
 
The effect of interference between resonant signal production and non-resonant
background diphoton production is studied using samples of simulated
diphoton events including contributions from both processes (produced by
either the $gg$ or $qg$ partonic channels) and their
interference. The accuracy of the calculations is NLO for the
$gg$-interference and LO for the $qg$-interference samples.
The events were generated using \Sherpa[2.2.11] and passed
through the \GEANT detector simulation.

% End of text imported from the .//sections/03-samples.tex input file

% The next lines are included from the .//sections/04-selection.tex input file
\section{Event selection and classification}
\label{sec:selection}
 
The event reconstruction and selection closely resemble those
used in the latest ATLAS measurement of Higgs boson production
cross-sections using the diphoton channel and the full \RunTwo
dataset~\cite{ATLAS:2022tnm}.
The main differences are the use of an updated photon energy
calibration~\cite{newegcalib} with reduced systematic uncertainties,
and the classification of events into categories that
are optimised to minimise the uncertainty in the measured Higgs boson mass
rather than in the Higgs boson production cross-sections.
 
Compared with the previous mass measurement~\cite{HIGG-2016-33}, which
used photon candidates reconstructed from a fixed-size
cluster of energy deposits in the electromagnetic calorimeter identified
by a sliding-window algorithm, this measurement relies on photon candidates
that are reconstructed from dynamic,
variable-size clusters, called superclusters.
The main advantages of this algorithm are improvements in the reconstruction
efficiency and energy measurement of converted photons ($\gamma \to e^+ e^-$),
and a lower rate of
misclassifying unconverted photons as converted photon
candidates~\cite{EGAM-2018-01}.
 
The photon energy is determined from the signals due to energy deposited in the
electromagnetic calorimeter, after applying the calibration
scheme detailed in Ref.~\cite{newegcalib}.
The photon direction is calculated from the
positions of the supercluster and the $pp$ collision vertex that is
chosen from among the reconstructed primary vertex candidates by a
neural-network (NN) algorithm~\cite{HIGG-2013-08}.
The NN inputs are the directions of the two $\pT$-leading photon candidates
in the event, determined only from the conversion vertices
and the longitudinal sampling of the calorimeter,
and vertex
candidate information such as the transverse momenta and directions
of the associated tracks.
 
The selection retains events with at least two photon candidates with
pseudorapidity in the range $|\eta|<1.37$ or $1.52<|\eta|<2.37$, meeting
{\em tight} identification and {\em loose} isolation
criteria~\cite{EGAM-2018-01}, matched to the
online photon candidates that passed the trigger selection.
Events are kept if the \pT-leading and \pT-subleading photon candidates
have invariant mass \myy in the range 105--160~\GeV\ and transverse
momenta that exceed 0.35 and 0.25 times \myy,
respectively.
When more than two photon candidates pass those requirements,
only the \pT-leading and \pT-subleading candidates are considered
for further analysis.
 
About 1.2~million events in the data pass the selection.
The expected efficiency for the signal for $m_H=125$~\GeV\
is close to 36\%,
leading to an expected signal yield of about 6200 events.
 
To increase the precision of the mass measurement, the selected
events are classified into 14 categories with different signal-to-background
ratios, diphoton invariant mass resolutions and photon energy scale
uncertainties.
The observables and the thresholds used to define the categories are
optimised by minimising the expected total Higgs boson mass
uncertainty for $m_H=125.09$~\GeV\
using a simplified version of the maximum-likelihood fit
described in the next section, including the
statistical uncertainties and the dominant systematic uncertainties
from the photon energy scale calibration.
The final choice of observables and thresholds for the categories used
in the measurement is the following:
\begin{itemize}
\item The number of reconstructed converted photon candidates:
events with no photon conversion candidates
(\enquote{U}-type events)
are considered separately from events with one or two $\gamma\to e^+e^-$
candidates (\enquote{C}-type events).
\item The absolute value of the pseudorapidity $|\eta_{S2}|$ of each of the
two energy clusters reconstructed in the electromagnetic calorimeter and
associated with the photon candidates.
The pseudorapidity $\eta_{S2}$ is determined
from the position of the barycentre of the cluster in the second sampling
layer of the calorimeter and from the
origin of the ATLAS coordinate system.
Both the U-type and C-type events are separated into three subsamples:
\enquote{central barrel} (both photons have  $|\eta_{S2}| < 0.8$),
\enquote{outer-barrel} (both photons have  $|\eta_{S2}| < 1.37$ and
at least one of these has $|\eta_{S2}| \geq 0.8$),
and \enquote{endcap} (at least one photon has
$1.52 \leq |\eta_{S2}| < 2.37$).
\item The magnitude $p^{\gamma\gamma}_\mathrm{Tt} = |\vec{p}_\mathrm{T}^{\gamma\gamma}\times \hat{t}|$ of the component of the diphoton transverse momentum
that is orthogonal to the thrust axis, defined as $\hat{t} = (\vec{p}_\mathrm{T}^{\gamma_1}-\vec{p}_\mathrm{T}^{\gamma_2})/|\vec{p}_\mathrm{T}^{\gamma_1}-\vec{p}_\mathrm{T}^{\gamma_2}|$.
Low ($p^{\gamma\gamma}_\mathrm{Tt}<70$~\GeV),
medium (70~\GeV$<p^{\gamma\gamma}_\mathrm{Tt}<130$~\GeV)
and high ($p^{\gamma\gamma}_\mathrm{Tt}>130$~\GeV) $p^{\gamma\gamma}_\mathrm{Tt}$
categories are defined for U-type and C-type central-barrel and
outer-barrel events.
\end{itemize}
 
For each category, the narrowest diphoton invariant mass window, with
half-width denoted by  $\sigma_{90}^{\gamma\gamma}$,
containing 90\% of the signal events is listed in
Table~\ref{tab:s90_b90_table}.
The expected signal ($S_{90}$) and background ($B_{90}$) yields in each category
in that interval are also reported, where $B_{90}$ is determined from the
integral of an exponentiated second-order polynomial function
fitted to the data $m_{\gamma\gamma}$ distribution after excluding the
range $120<m_{\gamma\gamma}<130$~\GeV\ from the interval.
The table also indicates the expected
fraction of signal events $f_{90} = S_{90}/(S_{90}+B_{90})$,
and the signal significance
$Z_{90} = \sqrt{\,2\left[(S_{90}+B_{90})\ln\left(1+S_{90}/B_{90}\right)-S_{90}\right]}$~\cite{Cowan:2010js}.
 
\begin{table}
\centering
\caption{The expected signal ($S_{90}$) and background ($B_{90}$)
yields, the signal yield as a percentage of the total ($f_{90}$),
and the signal significance ($Z_{90}$) in a diphoton invariant mass window
whose half-width $\sigma_{90}^{\gamma\gamma}$ is chosen in such a way that it
is the narrowest interval containing 90\% of signal events.
All quantities are given for each analysis
category and for the inclusive case.
}
\label{tab:s90_b90_table}
\begin{tabular}{lrrrrr}
\toprule
Category  & $\sigma_{90}^{\yy} [\si{\GeV}]$ & $S_{90}$ & $B_{90}$ & $f_{90}$ [\%] &  $Z_{90}$ \\
\midrule
U, Central-barrel, high \pTt   & 1.88 &    42 &    65 &   39.1 &  4.7     \\
U, Central-barrel, medium \pTt & 2.34 &   102 &   559 &   15.4 &  4.2     \\
U, Central-barrel, low \pTt    & 2.63 &   837 & 13\;\!226 &    6.0 &  7.2     \\
U, Outer-barrel, high \pTt     & 2.16 &    31 &    83 &   27.4 &  3.3     \\
U, Outer-barrel, medium \pTt   & 2.63 &   108 &   981 &    9.9 &  3.4     \\
U, Outer-barrel, low \pTt      & 3.00 &   869 & 22\;\!919 &    3.7 &  5.7     \\
U, Endcap                      & 3.33 &   759 & 29\;\!383 &    2.5 &  4.4     \\
C, Central-barrel, high \pTt   & 2.10 &    26 &    44 &   37.3 &  3.6     \\
C, Central-barrel, medium \pTt & 2.62 &    62 &   389 &   13.8 &  3.1     \\
C, Central-barrel, low \pTt    & 3.00 &   508 &  9\;\!726 &    5.0 &  5.1     \\
C, Outer-barrel, high \pTt     & 2.56 &    34 &   103 &   25.0 &  3.2     \\
C, Outer-barrel, medium \pTt   & 3.20 &   114 &  1\;\!353 &    7.8 &  3.1     \\
C, Outer-barrel, low \pTt      & 3.71 &   914 & 30\;\!121 &    2.9 &  5.2     \\
C, Endcap                      & 4.04 &  1\;\!249 & 52\;\!160 &    2.3 &  5.5     \\
\midrule
Inclusive                      & 3.32 &  5\;\!653 & 128\;\!774 &   4.2 & 15.6   \\
\bottomrule
\end{tabular}
\end{table}
 
The invariant mass resolution $\sigma_{90}^{\gamma\gamma}$
for C-type events is 10\%--20\% worse than for U-type events,
due to asymmetric $\gamma\to e^+e^-$ conversions
producing a low-energy electron or positron,
and to bremsstrahlung photons emitted by the
$e^+e^-$ pair. In both cases, such soft electrons, positrons or bremsstrahlung photons
can escape the supercluster and the calibration procedure may not be fully
efficient in recovering the energy of the original photon.
The resolution is 6\%--20\% better in the central-barrel categories than in the
corresponding outer-barrel categories, due to the smaller amount
of material upstream of the electromagnetic calorimeter
in the central region of the detector.
Events with high $p^{\gamma\gamma}_\mathrm{Tt}$ have about 15\%--20\% (30\%) better
$\myy$ resolution than events with medium (low) $p^{\gamma\gamma}_\mathrm{Tt}$,
due to poorer photon energy resolution at lower photon transverse momentum.
The signal fraction $f_{90}$ also depends on the same quantities:
it is larger for U-type events and central-barrel events, and increases with
$p^{\gamma\gamma}_\mathrm{Tt}$ since the main background process,
continuum diphoton production arising mostly from $t$-channel
$q\bar{q}\to\gamma\gamma$ and $gg\to\gamma\gamma$ scattering,
has a softer $p^{\gamma\gamma}_\mathrm{Tt}$ spectrum than the signal.
The photon energy scale uncertainty is smaller for C-type events and
central-barrel events than for U-type events and outer-barrel or endcap events;
it increases with $p^{\gamma\gamma}_\mathrm{Tt}$ due to uncertainties
in the linearity of the response and in the extrapolation to photons
in the energy scale calibration, which is mainly determined using
electron and positron candidates with relatively low transverse momentum from
$Z\to e^+ e^-$ decays.
 
The chosen categorisation reduces
the expected uncertainty in $m_H$ from statistical and
photon energy scale systematic uncertainties by about 17\% compared with
a measurement based on the inclusive sample passing the event
selection, and by 6\% compared with the use of the
event classification strategy in Ref.~\cite{HIGG-2016-33}.
Compared with the 101 event categories developed in
Ref.~\cite{ATLAS:2022tnm} for the measurement of the
Higgs boson production-mode cross-sections times branching ratio
to diphotons using the same dataset, the 14 categories used in this
analysis
lead to a  small increase ($+3\%$) in the expected statistical uncertainty
and to a larger decrease ($-14\%$) in the systematic uncertainty of $m_H$,
yielding an overall expected decrease ($-3\%$) in the total uncertainty.

% End of text imported from the .//sections/04-selection.tex input file

% The next lines are included from the .//sections/05-methodology.tex input file
\section{Mass measurement procedure}
\label{sec:methodology}
 
The Higgs boson mass $m_H$ is measured using the statistical methods
described in Ref.~\cite{HIGG-2016-33}.
The diphoton invariant mass distribution of the data is used to define
a likelihood function $L$ depending on $m_H$ and on
additional parameters $\boldsymbol{\theta}$ describing the signal and
background normalisations, their $m_{\gamma\gamma}$ models, and
corresponding systematic uncertainties.
The profile likelihood ratio~\cite{HIGG-2012-17,Cowan:2010js} is then:
\begin{equation}
\Lambda(m_H) =
\frac{L\big(m_H,\hat{\hat{\boldsymbol{\theta}}}(m_H)\big)}{L\big(\hat{m}_H,\hat{\boldsymbol{\theta}}\big)},
\label{eq:PLL}
\end{equation}
where $\hat{\boldsymbol{\theta}}$ and $\hat{m}_{H}$ denote the
values of the parameters that maximise the likelihood function $L$,
while $\hat{\hat{\boldsymbol{\theta}}}(m_H)$ represent the values of the
parameters $\boldsymbol{\theta}$ that maximise $L$ for a given value of
the parameter $m_H$.
A numerical fit procedure determines the central value
$\hat{m}_H$ of the measurement and its 68\%
confidence interval, defined by all values of $m_H$ for
which $-2\ln\Lambda(m_H)<1$.
The fit uses the event counts in 25-\MeV-wide bins of the $\myy$
distribution in each category.
 
The likelihood function $L$ is computed from the product of
individual likelihood functions for the observed diphoton invariant
mass spectra in each category and distributions representing auxiliary
measurements that constrain the nuisance parameters associated with
the systematic uncertainties.
The model assumes that the observed
distribution arises from the sum of a signal and a background component,
whose shape and normalisation are inferred from the data
themselves, with some input from the simulation.
 
The \myy distribution of the simulated signal in each category is found
to be properly described, for any value of the Higgs boson mass in the
range [110,\,140]~\GeV, by a double-sided Crystal~Ball~\cite{Oreglia:1980cs}
probability density function, i.e.\ a function with a Gaussian core and
power-law tails.
The parameters ($\alpha_\pm$, $n_\pm$) describing the tails of the model do
not depend on $m_H$, while the parameters $\mu_\mathrm{CB}$ and
$\sigma_\mathrm{CB}$ describing the peak position and the resolution of the
core Gaussian component scale linearly with $m_H$,
$\mu_\mathrm{CB} = m_H + a_\mu + b_\mu (m_H-125~\GeV)$
and
$\sigma_\mathrm{CB} = a_\sigma + b_\sigma (m_H-125~\GeV)$.
The nominal values of the parameters $a_\mu$, $b_\mu$,
$a_\sigma$, $b_\sigma$, $\alpha_\pm$ and $n_\pm$ in each category
are determined by a simultaneous fit to the simulated diphoton invariant
mass spectra in that category for different $m_H$ hypotheses.
As a cross-check, the fit
is repeated after removing the $m_H=125$~\GeV\ distribution from the input
and new values of the $a_\mu$, $b_\mu$, $a_\sigma$, $b_\sigma$,
$\alpha_\pm$ and $n_\pm$ parameters are obtained.
The signal model for $m_H=125$~\GeV\ predicted by this parameterisation
is then compared with the model determined from a single fit, with
a Crystal-Ball lineshape with floating  $\mu_\mathrm{CB}$,
$\sigma_\mathrm{CB}$, $\alpha_\pm$ and $n_\pm$ parameters,
to the $m_H=125$~\GeV\ simulated signal events.
Good agreement is observed.
The nominal signal models for the categories with the
best and worst expected resolutions for a Higgs boson mass $m_H=125$~\GeV\
are shown in Figure~\ref{fig:myy_signal}.
 
\begin{figure}[!htbp]
\centering
\includegraphics[width=0.8\textwidth]{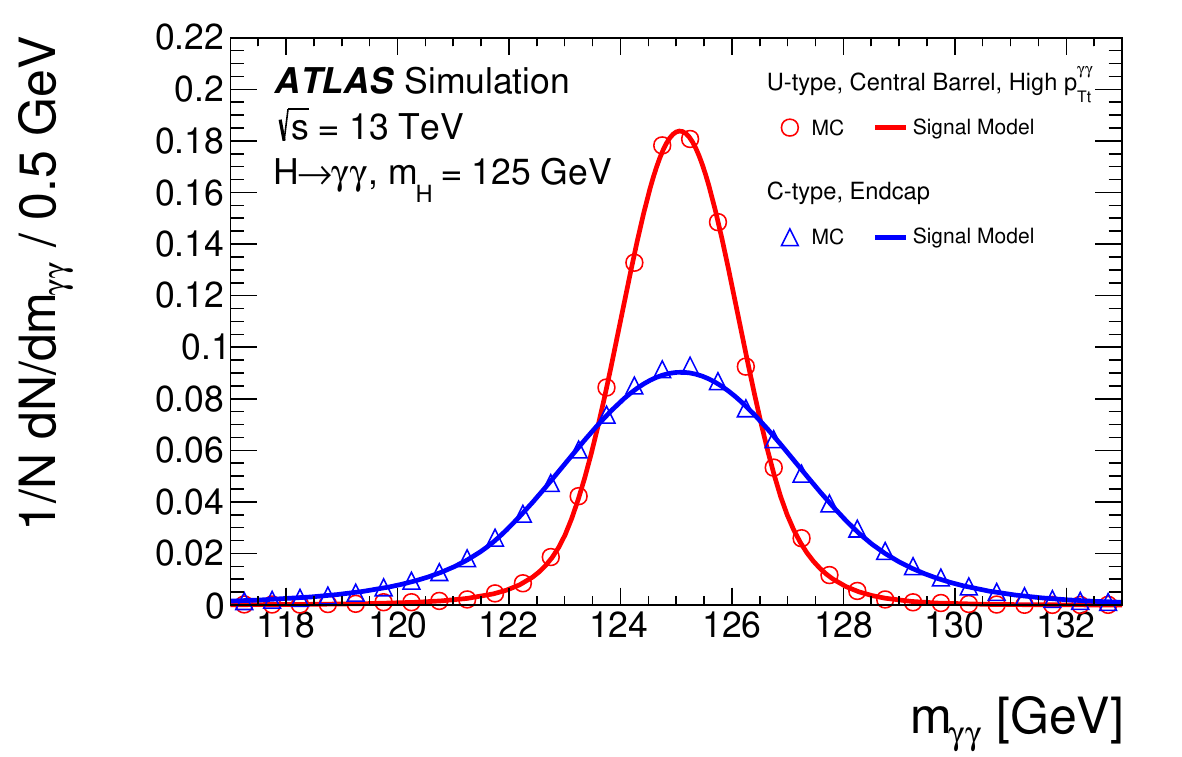}
\caption{Invariant mass distributions of simulated
$H\to\gamma\gamma$ events reconstructed in the categories with the best
(U-type, central-barrel, high-\pTt: open circles)
and the worst (C-type, endcap: open squares)
experimental resolutions.
The signal model derived from a fit of the simulated events is
superimposed (solid lines).
}
\label{fig:myy_signal}
\end{figure}
 
The normalisations of the signal, one for each category $c$, are free
parameters of the fit and are expressed as the product of a
per-category signal-strength factor $\mu_c$ and the expected number
$S_c$ of Higgs boson events in the same analysis region.
The expected signal yield $S_c$ is determined from the integrated luminosity,
the SM values of the Higgs boson cross-section and $H\to\gamma\gamma$ branching
ratio, and the simulation-predicted event selection efficiency in that category.
The dependence of $S_c$ on $m_H$ in each category $c$
is modelled with a second-order polynomial whose parameters are determined
by a fit to the expected yields in that category, calculated for nine
discrete values of $m_H$ between 110 and 140~\GeV\ using the simulated signal
samples described in Section~\ref{sec:samples}.
 
The background \myy distribution in each category is represented by either
an exponential function, a power-law function or an exponentiated second-order
polynomial.
The background model is chosen in an empirical way~\cite{ATLAS:2022tnm},
using the results obtained by fitting the diphoton invariant mass
distribution of a background template with models with free
parameters for the signal and background yields.
The background template is obtained by summing the \myy distributions
of simulated non-resonant diphoton events and data samples enriched
in photon+jet and dijet events.
The photon+jet and dijet enriched samples are obtained using a selection
similar to that in the previous section, the exception being that one or
both photon candidates must fail the nominal identification
and isolation requirements while passing looser ones.
The three distributions are normalised to the yields of the respective
contributions estimated in situ.
Among all considered background models whose $\chi^2$ probability
is greater than 1\% when fitted to the background template,
and for which the fitted signal yield (the \enquote{spurious signal}) is
less than 10\% of the expected signal yield,
the one with the fewest degrees of freedom is selected.
A value of 10\% is chosen for the threshold so that the spurious
signal, considered as a systematic uncertainty in the signal yield, is small
compared to the statistical uncertainty.
The background yield in each category and the parameters describing the
shape of the background model are free parameters of the likelihood
function.
 
Systematic uncertainties and their correlations are modelled by
including, among the parameters $\boldsymbol{\theta}$, nuisance
parameters described by likelihood functions associated with
estimates of the corresponding effects.
The statistical uncertainty of $m_H$ is estimated by determining the
confidence interval
when all nuisance parameters associated with systematic uncertainties
are fixed to their best-fit values, and all other
parameters are left unconstrained.
The
total systematic uncertainty is estimated as the square root of the
difference of the squares of the total uncertainty and the statistical uncertainty.
Similarly, the contribution from a group of uncertainties is determined from
the difference of the squares of the total uncertainty and the uncertainty
from a fit in which the associated nuisance parameters are fixed to their
best-fit values.
 
The expected statistical uncertainty of 120~\MeV\ in $m_H$ is a factor of 2.1
less than in the previous measurement~\cite{HIGG-2016-33},
due to the present dataset being almost four times larger and improvements
in the photon reconstruction algorithm and event classification.

% End of text imported from the .//sections/05-methodology.tex input file

% The next lines are included from the .//sections/06-syst.tex input file
\section{Systematic uncertainties}
\label{sec:syst}
 
The main sources of systematic uncertainty in \mH are the uncertainties
in the photon energy scale,
the uncertainty from the background modelling and the effect of
interference between the signal and the \yy continuum background.
They are described below, together with their expected pre-fit impact
on $m_H$.
 
The photon energy scale uncertainty is modelled by using 67 independent
components.
The effect of every component is evaluated for each analysis category
by comparing the nominal signal MC diphoton invariant mass distribution
with the one obtained by varying
the energy of each photon by the uncertainty under study.
The shifts induced in the peak parameters $\mu_\mathrm{CB}$ of the Crystal~Ball
signal models are then included as nuisance parameters in the likelihood
function, fully correlated among the categories.
The 67 independent sources of uncertainty in the photon energy scale
can be classified roughly into four main groups.
The first group (\textit{$\Zee$ calibration})
is related to the determination of the $\eta$-dependent energy scale
factors for electrons and positrons from $Z$ boson decays,
effectively constraining their energy scale for a transverse energy
$\et \approx 45$~\GeV.
The second group (\ET-\textit{dependent electron energy scale})
includes uncertainties
in the \et-dependence of the energy scale from sources such as
the calorimeter readout non-linearity, the calorimeter layer
intercalibration, and the amount of material upstream of the
calorimeter.
The third group (\textit{$e^\pm\rightarrow\gamma$ extrapolation})
includes the uncertainties
in the extrapolation of the energy scale from electrons to photons,
arising, for instance, from potential mismodelling of differences in
lateral shower development between electrons and photons in the calorimeter.
Finally, the fourth group (\textit{conversion modelling})
contains the uncertainties related to the accuracy of the photon
conversion modelling in the simulation.
Since the simulation-based photon energy calibration~\cite{newegcalib}
is trained and applied separately for unconverted and converted photon
candidates, any mismodelling of the conversion reconstruction performance
in the simulation may affect the calibrated photon energy scale.
 
The results presented in this Letter profit from a new auxiliary
measurement (\textit{linearity fit}) of the data-to-MC
electron energy scale corrections as a function of the electron \et.
In the photon energy scale calibration used in the previous
Higgs boson mass measurement~\cite{HIGG-2016-33}, only $\eta$-dependent
energy scale factors were derived from 2015--2016 data.
These were obtained by comparing the position
of the peak of the invariant mass distribution of $e^+e^-$ pairs
from $Z$ boson decays with that predicted by the simulation.
The possible \et-dependence of the data-to-MC energy scale correction
was accounted for as a systematic uncertainty, arising from the
various sources belonging to the \ET-dependent electron energy scale
group, calculated as described in Ref.~\cite{PERF-2017-03}.
The new approach used in Ref.~\cite{newegcalib} exploits the larger sample of
$Z\to e^+e^-$ decays collected in 2015--2018 to derive residual
data-to-MC energy scale factors in bins of electron transverse
energy within broad $\eta$ regions. The measurement of these additional
scale factors is used to constrain the
\et-dependent electron energy scale systematic uncertainties.
The additional constraints and correlation of the systematic uncertainties
from the linearity fit are propagated to the Higgs boson mass measurement
by implementing in the likelihood function a multivariate Gaussian
constraint term whose covariance is that returned by the
linearity fit.
The impact of the photon energy scale systematic uncertainties, including the
effect of the linearity fit, was found
to be independent of \mH for Higgs boson mass values in the range
124--126~\GeV, and is expected to be approximately $\pm 83$~\MeV\
(with contributions of 60, 43, 30 and 23~\MeV, respectively, from the
four individual groups of
$\Zee$ calibration, \ET-dependent electron energy scale,
$e^\pm\rightarrow\gamma$ extrapolation and conversion modelling
uncertainties).
Compared to the results in Ref.~\cite{HIGG-2016-33},
the simulation's more accurate description~\cite{EGAM-2018-01}
of material upstream of the EM calorimeter
and the lower sensitivity of the new clustering algorithm to the effects of
interactions with detector material
reduced the associated systematic uncertainty
by a factor close to three. In addition,
the larger dataset allowed a more precise study of the
$e^\pm\rightarrow\gamma$ extrapolation procedure~\cite{newegcalib},
and its impact on the expected \mH uncertainty is reduced by
a factor larger than three. Furthermore, the contribution of the
\ET-dependent electron energy scale uncertainties to the total
expected \mH uncertainty has been reduced by a factor of two thanks to more
precise dedicated measurements.
Finally, the linearity fit constrains the expected uncertainty
in the Higgs boson mass from the group of \ET-dependent energy scale
uncertainties by a further factor of four.
 
The effect of interference between the \ggHyy signal and the
\ggyy continuum background and between \gqHyyq and \gqyyq is not included
in the simulated event samples used to study the nominal signal model.
In the Standard Model, this interference is expected to change the
signal cross-section by 1\%--2\%~\cite{PhysRevLett.90.252001} and to
shift the position of the peak in the diphoton invariant mass distribution
by a few tens of \MeV~\cite{ATL-PHYS-PUB-2016-009}.
The size of this effect is treated as a systematic uncertainty,
which is quantified by fitting the nominal
signal-plus-background model to the sum of the \myy distributions
predicted by the nominal signal and background models and the \myy
distribution arising from such interference in simulated events.
The relative differences between the nominal Higgs boson mass and the
fitted mass in the various analysis categories are included in the likelihood
function as a single nuisance parameter affecting the 14 categories coherently.
The impact of the interference term on the determination of \mH is
expected to be approximately $\pm 24$~\MeV.
 
The effect of a possible bias in the measured \mH value, due to mismodelling
of the continuum background \myy distribution, is evaluated by fitting
the signal-plus-background model to the sum
of the \myy distribution of the background template described in the previous
section and the \myy distribution predicted by the signal model for
$\mH=125$~\GeV.
The relative differences between the nominal Higgs boson mass and the
fitted mass in the various analysis categories are then included in the
likelihood function as 14 additional uncorrelated nuisance parameters,
one per category.
The impact of the background modelling uncertainty on the measurement
of \mH is expected to be approximately $\pm 18$~\MeV.
 
The systematic uncertainty related to the selection of the diphoton
production vertex is evaluated in a \Zee control sample.
The directions of the selected electrons and positrons and thus their
invariant mass are calculated using either the primary vertex
candidate with the largest sum of the squared transverse momenta of
the tracks associated to the vertex (which includes also the $e^\pm$
tracks),
or the primary vertex
selected by the NN algorithm described in Section~\ref{sec:selection},
where the electron and positron tracks are ignored.
The separation between the peak positions of the two \ee invariant
mass distributions is evaluated separately in data and simulation.
The maximum difference (5~\MeV) between the separation observed in data
and that observed in the simulation is taken as an additional systematic
uncertainty in \mH.
 
Uncertainties from the chosen parameterisation of the nominal signal model
are propagated to the final result by including 14 additional uncorrelated
nuisance parameters in the likelihood function. They are determined
by fitting the nominal signal-plus-background model to the
sum of the \myy distribution of simulated $\mH=125$~\GeV\ signal events
and that predicted by the nominal background model.
The impact of the signal modelling uncertainty on the measurement of \mH
is expected to be approximately $\pm 5$~\MeV.
 
The effect of the photon energy resolution uncertainty is included in the
signal model as five nuisance parameters that affect the resolution parameter
$\sigma_\mathrm{CB}$ of the Crystal~Ball function and are treated as
being correlated among categories.
The impact of the five independent sources of photon energy resolution
uncertainty is evaluated by comparing the nominal signal invariant mass
distribution in each category with the ones obtained by varying
the energy resolution of each photon according to its uncertainties.
The sum in quadrature of the different components of the photon energy
resolution uncertainty ranges from 4.5\% for C-type, outer-barrel,
low-\pTt events to 17\% for U-type, central-barrel, high-\pTt events.
The impact of the photon energy resolution uncertainty on \mH
is expected to be approximately $\pm 3$~\MeV.
 
Yield uncertainties from the Higgs boson branching ratio to diphotons and
the integrated luminosity of the data, and uncertainties in the migrations
of events between categories, from various experimental and theoretical
sources, are included in the model.
Experimental uncertainties in
the efficiency of photon conversion reconstruction,
photon identification, photon isolation and the photon triggers,
as well as in the impact of the simulation's modelling of \pileup
are considered.
Theoretical uncertainties that are taken into account are
those in the signal production cross-sections,
in the modelling of the underlying event and parton shower,
in the value of the strong coupling constant and
in the parton distribution functions of the proton.
All the uncertainties described in this paragraph are included in the fit,
although their expected impact on the measurement of \mH was found
to be below 1~\MeV.
 
In summary, compared to the previous publication~\cite{HIGG-2016-33}
based on 36~\ifb of ATLAS \RunTwo data, the improvements in the
photon energy calibration~\cite{newegcalib}
and the optimised event classification lead to an expected fourfold reduction
of the dominant systematic uncertainty
in $m_H$ (from the photon energy scale and resolution),
from 320~\MeV\ to 80~\MeV,
and a similar reduction of the total expected systematic uncertainty,
from 330~\MeV to 90~\MeV.

% End of text imported from the .//sections/06-syst.tex input file

% The next lines are included from the .//sections/07-results.tex input file
\section{Results}
\label{sec:results}

The \myy\ distribution of the data, overlaid with the sum of the
signal and background models corresponding to the maximum-likelihood
estimates of the parameters of the likelihood function, is shown
in Figure~\ref{fig:myy_data_w}. All event categories are
included. For illustration purposes, events
in each category are weighted by a factor $\ln(1 + S^{\mathrm{obs}}_{90}/B^{\mathrm{obs}}_{90})$, where
$S^{\mathrm{obs}}_{90}$ and $B^{\mathrm{obs}}_{90}$ are the fitted signal and background yields in the smallest
$m_{\gamma\gamma}$ interval containing 90\% of the signal.
\begin{figure}[!htbp]
\centering
\includegraphics[width=0.8\textwidth]{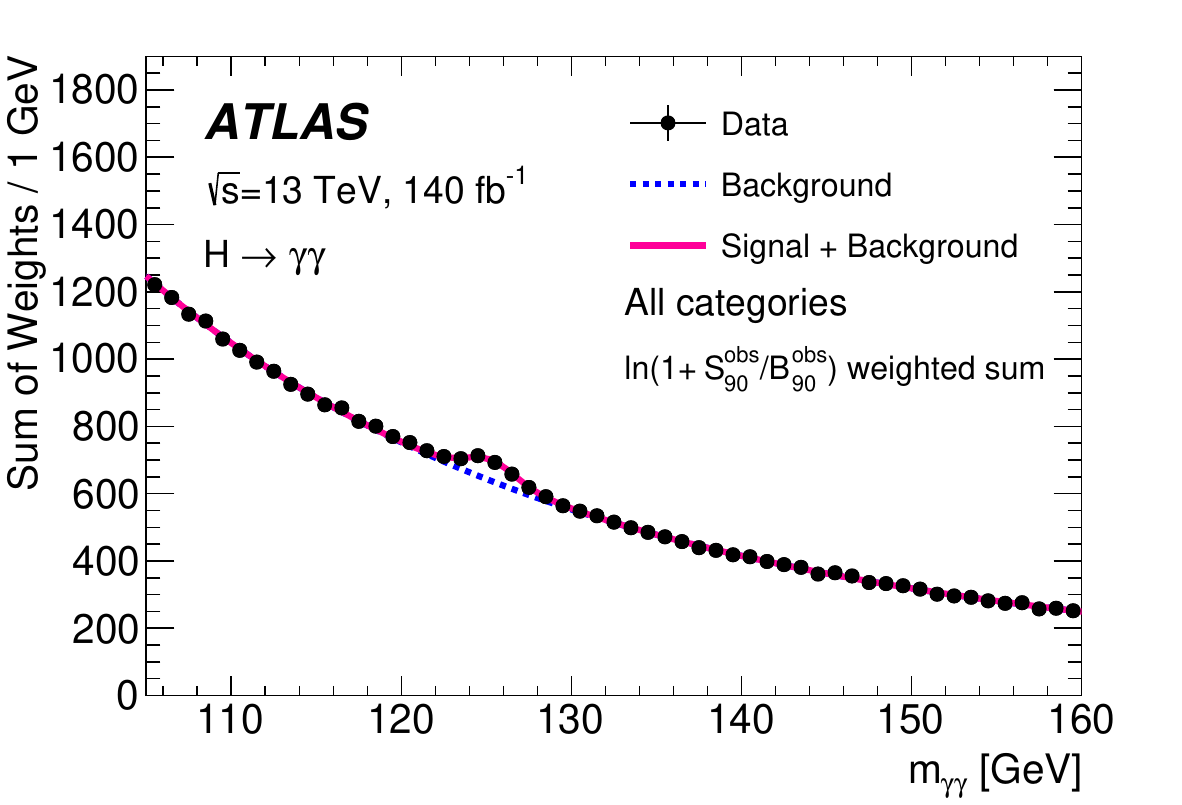}
\caption{Diphoton invariant mass
distribution of all selected data events (black dots with error bars),
overlaid with the result of the fit (solid red line).
For both the data and the fit, each category is weighted by a
factor $\ln(1 + S^{\mathrm{obs}}_{90}/B^{\mathrm{obs}}_{90})$,
where $S^{\mathrm{obs}}_{90}$ and $B^{\mathrm{obs}}_{90}$ are the fitted signal
and background yields in the smallest $m_{\gamma\gamma}$ interval containing
90\% of the expected signal. The dotted line describes the background
component of the model.
}
\label{fig:myy_data_w}
\end{figure}
 
The profile likelihood ratio as a function of $m_H$
is shown in Figure~\ref{fig:mH_lh_scan}.
The value of the Higgs boson mass determined by fitting the profile
likelihood ratio in Eq.~(\ref{eq:PLL}) to the diphoton invariant mass
distribution in data is:
\begin{equation*}
m_H = 125.17 \pm 0.11 (\mathrm{stat.}) \pm 0.09 (\mathrm{syst.})~\GeV = 125.17 \pm 0.14~\GeV.
\end{equation*}
The statistical and systematic uncertainties are
in good agreement
with the values of 120~\MeV\ and 90~\MeV\ expected for a SM Higgs
boson with the observed mass.
The main sources of systematic uncertainty and their impact on the measurement
are summarised in Table~\ref{tab:yy_sys_postfit}.
 
The signal strength $\mu_c$ in each category $c$ is compatible with the
SM prediction $\mu_c=1$, with the largest difference being 2.2
standard deviations ($\sigma$)
for the C-type, central-barrel, medium-\pTt\ category.
The global significance of this difference, taking into account
a trial factor of 14, is less than $1\sigma$.
The best-fit $m_H$ values for the individual categories
are in good agreement with each other, with a global $p$-value
of 8\%.
If the same signal strength $\mu$ is used for each category,
the central value of $m_H$ is shifted by $-35$~\MeV, and the fitted value of
$\mu$ is in agreement with the SM prediction within 1.4 standard deviations.
If the signal $\myy$ model is modified to account for the expected
shift induced by interference with non-resonant background
diphoton production,
the measured value of the Higgs boson mass is increased by
approximately 26~\MeV.
 
\begin{table}[htbp]
\centering
\caption{
Estimated impact of the main sources of systematic uncertainty
on the $m_H$ measurement with \RunTwo data.
}
\label{tab:yy_sys_postfit}
\begin{tabular}{lr}
\toprule
Source & Impact [MeV]\\
\midrule
Photon energy scale                                  & $ 83$ \\
~~~~$\Zee$ calibration                               & $ 59$ \\
~~~~$\ET$-dependent electron energy scale            & $ 44$ \\
~~~~$e^\pm\rightarrow\gamma$ extrapolation           & $ 30$ \\
~~~~Conversion modelling                             & $ 24$ \\
Signal--background interference                       & $ 26$ \\
Resolution                                           & $ 15$ \\
Background model                                     & $ 14$ \\
Selection of the diphoton production vertex          & $ 5$ \\
Signal model                                         & $ 1$ \\
\midrule
Total                                                & $ 90$ \\
\bottomrule
\end{tabular}
\end{table}
 
\begin{figure}[!htbp]
\centering
\begin{tabular}{@{}c@{}}
\subfloat[]{
\includegraphics[width=0.46\textwidth]{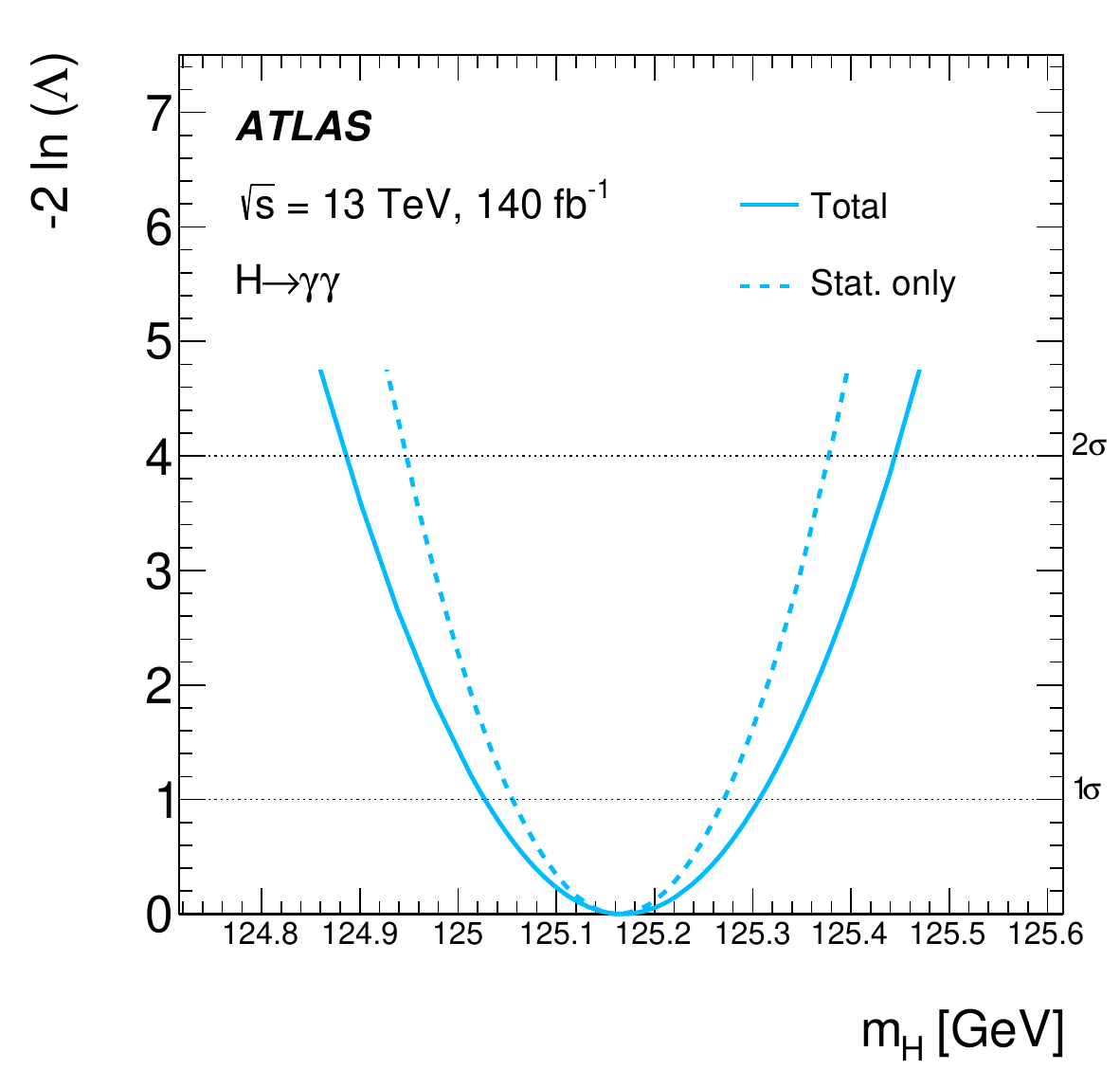}
\label{fig:mH_lh_scan}
}
\end{tabular}
\begin{tabular}{@{}c@{}}
\subfloat[]{
\includegraphics[width=0.46\textwidth]{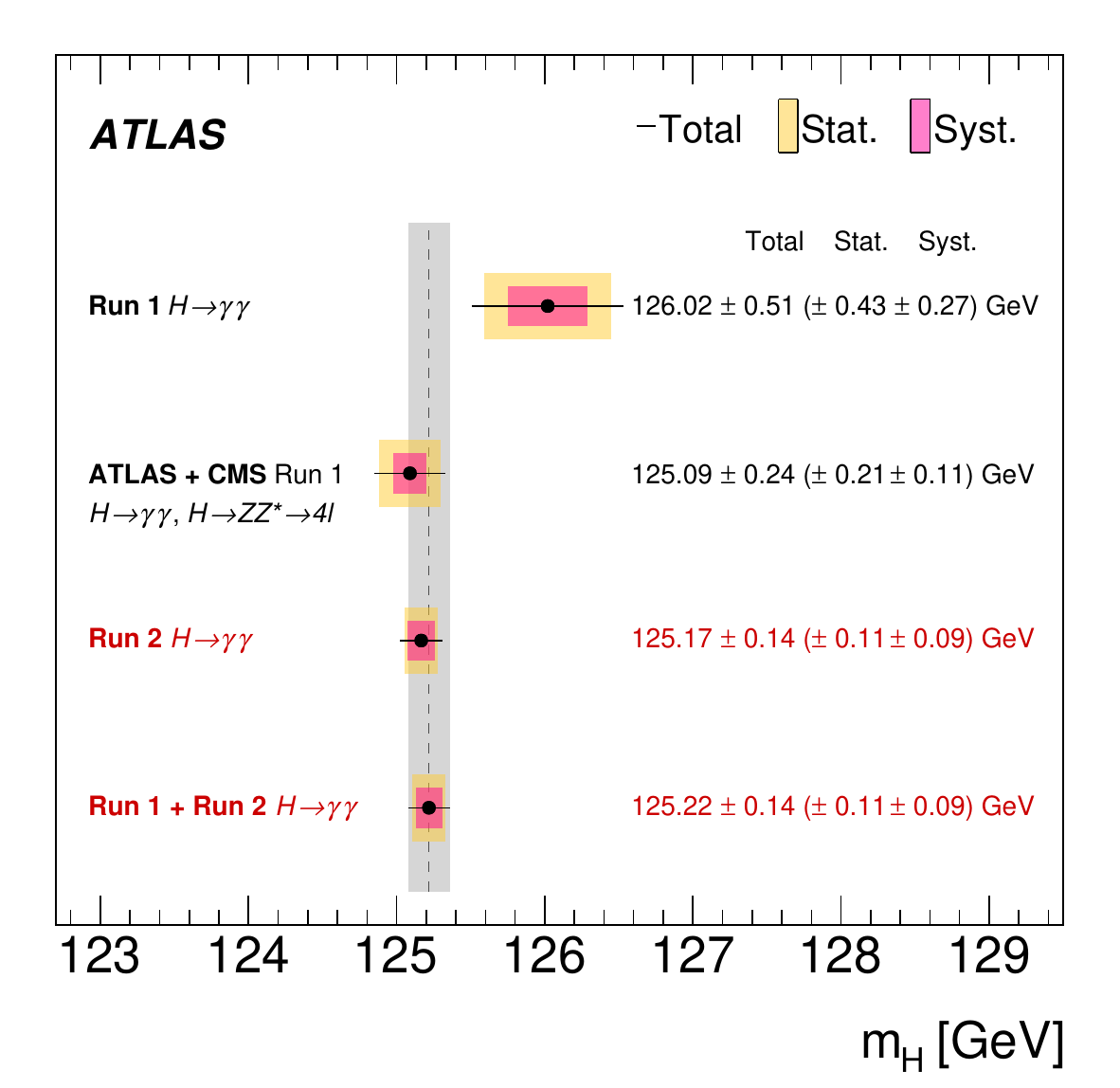}
\label{fig:mH_comb}
}
\end{tabular}
\caption{
\protect\subref{fig:mH_lh_scan}
Value of $-2\ln\Lambda$ as a function of $m_H$
for the combined fit to all $H\to\gamma\gamma$ categories.
The intersections of the $-2\ln\Lambda$ curve with the horizontal
lines labelled $1\sigma$ and $2\sigma$ provide the 68.3\%
and 95.5\% confidence intervals.
\protect\subref{fig:mH_comb}
Summary of the Higgs boson mass measurements from the
analysis of $H\to\gamma\gamma$ decays in ATLAS \RunTwo data and
combined \RunOne~+~\RunTwo data presented in this
Letter,
compared with the combined \RunOne ATLAS
result in the diphoton channel and with the \RunOne
measurement by ATLAS and CMS~\cite{HIGG-2014-14} combining the diphoton
and four-lepton channels.
The statistical, systematic and total uncertainties are indicated
with horizontal yellow-shaded bands, pink-shaded bands and black error
bars, respectively.
The vertical dashed line and grey shaded band around it indicate the central
value and total uncertainty of the $H\to\gamma\gamma$ ATLAS \RunOne~+~\RunTwo
measurement.
}
\end{figure}
 
The present measurement is combined with the previous one,
$m_H = 126.02 \pm 0.43 \mathrm{(stat.)} \pm 0.27 \mathrm{(syst.)}$~\GeV,
obtained by ATLAS in the diphoton channel using $25$~\ifb~of proton--proton
collisions recorded at $\sqrt{s}=7$ and 8~\TeV\ during
\RunOne of the LHC in 2011--2012~\cite{HIGG-2014-14}.
The combination of ATLAS \RunOne and \RunTwo results is performed by
simultaneously fitting a single $m_H$ parameter to the two datasets.
The nominal model including the 14 signal strengths $\mu_c$ of the
reconstructed categories is used for the \RunTwo dataset, while two separate
signal strengths, one each for production processes involving
Higgs boson couplings to either fermions or vector bosons, are used for the
\RunOne dataset. All 16 signal-strength parameters are
profiled in the combined fit for $m_H$.
Almost all the nuisance parameters, especially the
\et-dependent photon energy scale parameters affected by the linearity
measurement and the parameters describing the extrapolation of the
energy scale from electrons to photons, are assumed to be uncorrelated
between the two measurements because of differences in the reconstruction
algorithms and in the calibration procedures and control samples.
The $\Zee$ scale uncertainties and some of the resolution
uncertainties, estimated in the same way for the two measurements,
are treated as being fully correlated between the two data-taking periods.
The combination with the \RunOne ATLAS measurement in the diphoton channel
produces a small shift ($+50$~\MeV) of the
central value and a slight reduction (${<}10$~\MeV) of the statistical
uncertainty, leading to:
\begin{equation*}
m_H = 125.22 \pm 0.11 (\mathrm{stat.}) \pm 0.09 (\mathrm{syst.})~\GeV = 125.22 \pm 0.14~\GeV.
\end{equation*}
 
The individual ATLAS \RunOne and \RunTwo measurements in the diphoton channel
and their combination are shown in Figure~\ref{fig:mH_comb}, together with the ATLAS+CMS
\RunOne measurement~\cite{HIGG-2014-14} using $H\to\gamma\gamma$ and $H\to 4\ell$ decays.

% End of text imported from the .//sections/07-results.tex input file

\FloatBarrier

% The next lines are included from the .//sections/08-conclusion.tex input file
\section{Conclusion}
\label{sec:conclusion}
 
A measurement of the Higgs boson mass in the diphoton channel has been
performed using the \RunTwo $pp$ collision data recorded by the ATLAS
experiment at the CERN Large Hadron Collider at a centre-of-mass energy of
13~\TeV, corresponding to an integrated luminosity of 140~fb$^{-1}$.
Compared to a previous measurement in the same decay channel,
which was based on a four times smaller dataset at the same energy,
this measurement has a systematic (total) uncertainty that is reduced
by a factor close to four (three).
The reduction in systematic uncertainty is mainly due to an improved
photon energy scale calibration, with better energy resolution, and smaller
\ET-dependent and $e^\pm \to \gamma$ extrapolation uncertainties.
 
The Higgs boson mass is measured to be $m_H=125.17\pm 0.14$~GeV.
Combined with the \RunOne ATLAS result in the same decay channel,
it yields a value $m_H = 125.22\pm 0.14$~\GeV.
With an uncertainty of 1.1 per mille, this is currently the most
precise measurement of the Higgs boson mass from a single channel.

% End of text imported from the .//sections/08-conclusion.tex input file

\section*{Acknowledgements}

% The next lines are included from the .//acknowledgements/Acknowledgements.tex input file

We thank CERN for the very successful operation of the LHC, as well as the
support staff from our institutions without whom ATLAS could not be
operated efficiently.
 
We acknowledge the support of
ANPCyT, Argentina;
YerPhI, Armenia;
ARC, Australia;
BMWFW and FWF, Austria;
ANAS, Azerbaijan;
CNPq and FAPESP, Brazil;
NSERC, NRC and CFI, Canada;
CERN;
ANID, Chile;
CAS, MOST and NSFC, China;
Minciencias, Colombia;
MEYS CR, Czech Republic;
DNRF and DNSRC, Denmark;
IN2P3-CNRS and CEA-DRF/IRFU, France;
SRNSFG, Georgia;
BMBF, HGF and MPG, Germany;
GSRI, Greece;
RGC and Hong Kong SAR, China;
ISF and Benoziyo Center, Israel;
INFN, Italy;
MEXT and JSPS, Japan;
CNRST, Morocco;
NWO, Netherlands;
RCN, Norway;
MEiN, Poland;
FCT, Portugal;
MNE/IFA, Romania;
MESTD, Serbia;
MSSR, Slovakia;
ARRS and MIZ\v{S}, Slovenia;
DSI/NRF, South Africa;
MICINN, Spain;
SRC and Wallenberg Foundation, Sweden;
SERI, SNSF and Cantons of Bern and Geneva, Switzerland;
MOST, Taiwan;
TENMAK, T\"urkiye;
STFC, United Kingdom;
DOE and NSF, United States of America.
In addition, individual groups and members have received support from
BCKDF, CANARIE, Compute Canada and CRC, Canada;
PRIMUS 21/SCI/017 and UNCE SCI/013, Czech Republic;
COST, ERC, ERDF, Horizon 2020, ICSC-NextGenerationEU and Marie Sk{\l}odowska-Curie Actions, European Union;
Investissements d'Avenir Labex, Investissements d'Avenir Idex and ANR, France;
DFG and AvH Foundation, Germany;
Herakleitos, Thales and Aristeia programmes co-financed by EU-ESF and the Greek NSRF, Greece;
BSF-NSF and MINERVA, Israel;
Norwegian Financial Mechanism 2014-2021, Norway;
NCN and NAWA, Poland;
La Caixa Banking Foundation, CERCA Programme Generalitat de Catalunya and PROMETEO and GenT Programmes Generalitat Valenciana, Spain;
G\"{o}ran Gustafssons Stiftelse, Sweden;
The Royal Society and Leverhulme Trust, United Kingdom.
 
The crucial computing support from all WLCG partners is acknowledged gratefully, in particular from CERN, the ATLAS Tier-1 facilities at TRIUMF (Canada), NDGF (Denmark, Norway, Sweden), CC-IN2P3 (France), KIT/GridKA (Germany), INFN-CNAF (Italy), NL-T1 (Netherlands), PIC (Spain), ASGC (Taiwan), RAL (UK) and BNL (USA), the Tier-2 facilities worldwide and large non-WLCG resource providers. Major contributors of computing resources are listed in Ref.~\cite{ATL-SOFT-PUB-2023-001}.

% End of text imported from the .//acknowledgements/Acknowledgements.tex input file

\printbibliography
 
\clearpage
% ATLAS Collaboration author list
% Reference date of HIGG-2019-16 is 2023-06-28
% Author list last updated on date 14-DEC-23
% Data extracted on 14-Dec-2023 for paper reference HIGG-2019-16
% at 5:29pm
 
\begin{flushleft}
\hypersetup{urlcolor=black}
{\Large The ATLAS Collaboration}

\bigskip

\AtlasOrcid[0000-0002-6665-4934]{G.~Aad}$^\textrm{\scriptsize 102}$,
\AtlasOrcid[0000-0002-5888-2734]{B.~Abbott}$^\textrm{\scriptsize 120}$,
\AtlasOrcid[0000-0002-1002-1652]{K.~Abeling}$^\textrm{\scriptsize 55}$,
\AtlasOrcid[0000-0001-5763-2760]{N.J.~Abicht}$^\textrm{\scriptsize 49}$,
\AtlasOrcid[0000-0002-8496-9294]{S.H.~Abidi}$^\textrm{\scriptsize 29}$,
\AtlasOrcid[0000-0002-9987-2292]{A.~Aboulhorma}$^\textrm{\scriptsize 35e}$,
\AtlasOrcid[0000-0001-5329-6640]{H.~Abramowicz}$^\textrm{\scriptsize 151}$,
\AtlasOrcid[0000-0002-1599-2896]{H.~Abreu}$^\textrm{\scriptsize 150}$,
\AtlasOrcid[0000-0003-0403-3697]{Y.~Abulaiti}$^\textrm{\scriptsize 117}$,
\AtlasOrcid[0000-0002-8588-9157]{B.S.~Acharya}$^\textrm{\scriptsize 69a,69b,m}$
% ...
\bigskip
\\

% ...
$^{29}$Physics Department, Brookhaven National Laboratory, Upton NY; United States of America.\\
% ...
$^{35}$$^{(a)}$Facult\'e des Sciences Ain Chock, R\'eseau Universitaire de Physique des Hautes Energies - Universit\'e Hassan II, Casablanca;$^{(b)}$Facult\'{e} des Sciences, Universit\'{e} Ibn-Tofail, K\'{e}nitra;$^{(c)}$Facult\'e des Sciences Semlalia, Universit\'e Cadi Ayyad, LPHEA-Marrakech;$^{(d)}$LPMR, Facult\'e des Sciences, Universit\'e Mohamed Premier, Oujda;$^{(e)}$Facult\'e des sciences, Universit\'e Mohammed V, Rabat;$^{(f)}$Institute of Applied Physics, Mohammed VI Polytechnic University, Ben Guerir; Morocco.\\
% ...
$^{49}$Fakult\"{a}t Physik , Technische Universit{\"a}t Dortmund, Dortmund; Germany.\\
% ...
$^{55}$II. Physikalisches Institut, Georg-August-Universit\"{a}t G\"ottingen, G\"ottingen; Germany.\\
% ...
$^{69}$$^{(a)}$INFN Gruppo Collegato di Udine, Sezione di Trieste, Udine;$^{(b)}$ICTP, Trieste;$^{(c)}$Dipartimento Politecnico di Ingegneria e Architettura, Universit\`a di Udine, Udine; Italy.\\
% ...
$^{102}$CPPM, Aix-Marseille Universit\'e, CNRS/IN2P3, Marseille; France.\\
% ...
$^{117}$Department of Physics, New York University, New York NY; United States of America.\\
% ...
$^{120}$Homer L. Dodge Department of Physics and Astronomy, University of Oklahoma, Norman OK; United States of America.\\
% ...
$^{150}$Department of Physics, Technion, Israel Institute of Technology, Haifa; Israel.\\
$^{151}$Raymond and Beverly Sackler School of Physics and Astronomy, Tel Aviv University, Tel Aviv; Israel.\\
% ...
$^{a}$ Also Affiliated with an institute covered by a cooperation agreement with CERN.\\
$^{b}$ Also at An-Najah National University, Nablus; Palestine.\\
$^{c}$ Also at Borough of Manhattan Community College, City University of New York, New York NY; United States of America.\\
$^{d}$ Also at Center for High Energy Physics, Peking University; China.\\
$^{e}$ Also at Center for Interdisciplinary Research and Innovation (CIRI-AUTH), Thessaloniki; Greece.\\
$^{f}$ Also at Centro Studi e Ricerche Enrico Fermi; Italy.\\
% ...
$^{m}$ Also at Department of Physics, King's College London, London; United Kingdom.\\
% ...

\end{flushleft}

% Created with Glance <Atlas.Glance@cern.ch>

\end{document}